\documentclass[10pt,aps,pre,superscriptaddress,nofootinbib,twocolumn,12p]{revtex4-2}

\usepackage{amssymb}  
\usepackage{amsmath}  
\usepackage{bbold}
\usepackage{upgreek}
\usepackage{epsfig}   
\usepackage{MnSymbol}
\usepackage[english]{babel}
\usepackage{relsize}
\usepackage{graphicx,subfigure}   
\usepackage{dcolumn}
\usepackage{bm}  
\usepackage{soul}
\usepackage[T1]{fontenc}
\usepackage[colorlinks,citecolor=blue]{hyperref}
\RequirePackage{lineno}

\emergencystretch=\maxdimen
\newcommand{\SH}{\mathsf{SH}}
\newcommand{\RH}{\mathsf{RH}}
\newcommand{\SB}{\mathsf{SB}}
\newcommand{\RB}{\mathsf{RB}}

\newcommand{\SC}{\mathsf{SC}}
\newcommand{\RC}{\mathsf{RC}}

\newcommand{\SuppMat}{\emph{Supplementary Materials}}

\newcommand{\C}{\mathcal{C}}

\usepackage{etoolbox}
\usepackage{setspace}
\usepackage{mathtools}
\begin{document}

\title{Scaling of connectivity metrics in river networks}

\author{E. H.  Colombo} \email{e.colombo@hzdr.de}
\affiliation{Center for Advanced Systems Understanding (CASUS), Helmholtz-Zentrum Dresden Rossendorf (HZDR), Görlitz, Germany.}

\author{A. B. García-Andrade} 
\affiliation{Center for Advanced Systems Understanding (CASUS), Helmholtz-Zentrum Dresden Rossendorf (HZDR), Görlitz, Germany.}

\author{Ismail} 
\affiliation{Center for Advanced Systems Understanding (CASUS), Helmholtz-Zentrum Dresden Rossendorf (HZDR), Görlitz, Germany.}

\author{J. M. Calabrese} \email{j.calabrese@hzdr.de}
\affiliation{Center for Advanced Systems Understanding (CASUS), Helmholtz-Zentrum Dresden Rossendorf (HZDR), Görlitz, Germany.}
\affiliation{Department of Ecological Modelling, Helmholtz Centre for Environmental Research – UFZ, Leipzig, Germany}
\affiliation{Department of Biology, University of Maryland, College Park, MD, USA}


\noindent 
\begin{abstract}
Rivers exhibit fractal-like properties that are associated with scaling laws linking geometry and size. The optimal channel network (OCN) model, which is a mathematically tractable representation of river networks often used in theoretical studies, is based on the fractal properties of rivers and consequently reproduces geometric scaling laws. However, purely geometric relationships may not fully capture the interaction between river structure and species' movement strategies that is most relevant to many large-scale ecological processes. In contrast, connectivity, which is a concept that blends habitat geometry and individual movement, has been shown both theoretically and empirically to influence relevant large-scale ecological outcomes across a broad array of ecosystems. Here, we analyze networks from more than 1000 major rivers around the world, including the Amazon, Mississippi, and Nile, to investigate how river network connectivity metrics scale with system size. Specifically, we found clear power-law scaling of both the harmonic centrality and betweenness centrality network connectivity metrics. To assess the extent to which OCNs can capture these empirical connectivity patterns, we generated synthetic river networks by fitting an OCN model to each real river. We found excellent agreement between empirical and OCN-based scaling laws, supporting the notion that OCNs can accurately represent rivers in network-based models and analyses. Finally, we examined the robustness of the connectivity scaling laws to species movement strategies ranging from ideal shortest-path navigation to suboptimal random-path navigation. Surprisingly, we found that random navigation breaks the power-law scaling relationship for harmonic centrality, but not for betweenness centrality. 
\end{abstract}

\maketitle

\section{Introduction}
\label{sec:intro}

Rivers are well known to produce scale-invariant relationships between geometric features and measures of size, such as those connecting the longest stream length (Hack's law~\cite{hack1957studies}) and the basin slope (Flint's law~\cite{flint1974stream} to the basin area. In other words, geometric river data often collapse onto a power-law, $X\sim N^\beta$, where $X$ is a river feature, $N$ is a size-related variable, and $\beta$ is the scaling exponent. By demonstrating that rivers obey universal macroscopic laws, the work of Rodriguez-Iturbe, Rinaldo, and colleagues~\cite{rinaldo_minimum_1992, rodriguez1997fractal}, as well as recent extensions thereof~\cite{carraro_optimal_2022,wollheim2022superlinear}, has fundamentally changed our understanding of both the origin and function of rivers. In particular, this body of work has led to a general theory of river formation based on the physical principle of minimizing the total energy expenditure across a river basin~\cite{rinaldo_minimum_1992}.  Furthermore, this minimization process also provides a rigorous basis from which to derive so-called optimal channel networks (OCNs), which have been shown to reproduce many statistical properties of real river systems~\cite{riverocn}. Optimal channel networks are therefore frequently used as a mathematically tractable representation of rivers in theoretical studies~\cite{balister2018river}.

However, existing geometric scaling relationships do not necessarily capture the processes that are most crucial for understanding and predicting how ecological outcomes such as biodiversity or ecosystem services might change with system size. This deficiency stems from the fact that ecological processes often depend on the coupling between species traits and the structure of the environment. One prominent ecological concept that captures this kind of species-environment interaction is connectivity, which is an emergent property that arises from the interplay between habitat structure and species navigation strategies~\cite{calabrese2004comparison}. Connectivity quantifies how easily dispersing organisms can move from one spatial location to another. For strictly aquatic organisms (e.g., freshwater fish), connectivity depends strongly on the dendritic structure of river networks as well as species' abilities to navigate those networks. Additionally, as connectivity is known to influence many ecological outcomes (see examples below), it has the potential to serve as a better target for the study of ecologically relevant scaling relationships in riverine ecosystems.

%

\begin{figure*}[th]
    \centering
    \includegraphics[width=1\textwidth]{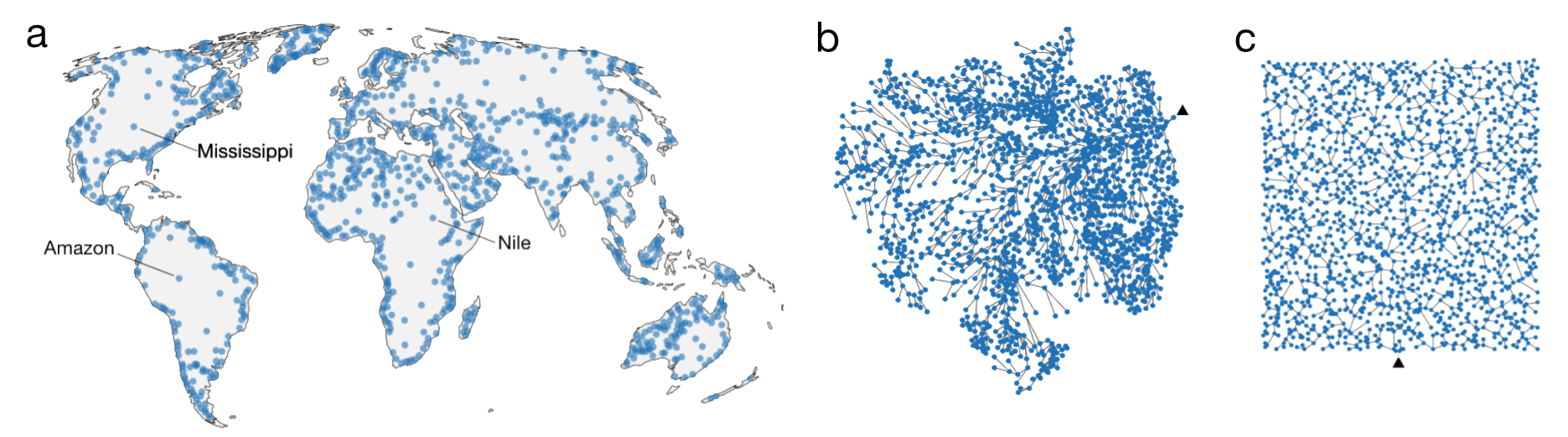}
    \caption{\textbf{Data summary and network extraction.} a) The total of 1139 river basins studied around the world (blue dots represent their centroids). b) The networks for the Amazon river and c) its OCN-derived counterpart which has same area and number of nodes. Black triangle on b) and c) indicate the outlet location. See Sec.~\ref{sec:methods} for details.}
    \label{fig:scheme}
\end{figure*}

Greater connectivity has been shown to correlate with higher species richness across a range of ecosystems including: i) Indo-Pacific coral reefs~\cite{cowen2009larval}; ii) river-connected vs isolated areas in the Amazon rain forest~\cite{Peres2005}; iii) corridor-connected vs isolated patches in the Atlantic Forest of Brazil ~\cite{haddad2015habitat}; iv) European grassland fragments varying in connectivity \cite{fahrig2003effects}; and v)  Marine Protected Areas in the Mediterranean Sea~\cite{roberts2001effects}. 
In parallel, a large body of theoretical work has also firmly established that connectivity promotes longer species persistence in metapopulations~\cite{hanski2000metapopulation,keymer2000extinction,gilarranz2012spatial,rocha2021dispersal,PhysRevE.92.022714}, leads to higher species diversity in metacommunities~\cite{macarthur2001theory,hanski1999metapopulation}, and can strongly influence the degree of species similarity between twolocations~\cite{economo_species_2008,economo_network_2010}. In the specific context of riverine communities, a simple neutral biodiversity model~\cite{hubbell2006neutral}, when modified to account for connectivity on a dendritic network, has been shown to accurately describe multi-scale patterns in fish biodiversity in the Mississippi-Missouri river system~\cite{muneepeerakul_neutral_2008}. Additionally, recent empirical and theoretical studies have established that connectivity is a critical link between the dendritic structure of rivers and ecological outcomes ranging from biodiversity generation and maintenance~\cite{mcintosh2024ecosystem,shao_conn_review_2019} to bio-geochemical processes and energy cycles~\cite{wollheim2022superlinear}.

Building on this abundant evidence supporting connectivity as a key driver of numerous ecological processes, we investigate the scaling of connectivity across 1139 river networks, including major basins such as the Amazon (Fig.~\ref{fig:scheme}b), Mississippi, Nile, and others.
We first assemble a dataset containing networks and then compute a diverse set of network connectivity indices including the well-known \mbox{harmonic-,} betweenness-, degree-, closeness-, and eigenvector-centrality metrics, and investigate their dependence on network size. Importantly, to clarify the role of species navigation strategies in shaping connectivity, we include indices based on the extreme assumptions of: optimal navigation, where pairs of nodes are always connected by the shortest possible path, and random navigation, where nodes are connected by random-walk paths. These extremes delimit the range of likely dispersal patterns that riverine fish species might exhibit in the wild. On the one hand, migratory species like salmon and sturgeons use accurate navigation mechanisms that lead to highly directed movement patterns~\cite{leggett1977ecology}. On the other hand, many fish species appear to disperse rather randomly across rivers, which is a fact validated by both empirical observations~\cite{radinger2014patterns} and model-based statistical analysis~\cite{muneepeerakul_neutral_2008}. We then created a synthetic river network dataset by fitting an optimal channel network (OCN) model to each real river, to ask if OCNs, which are widely used to represent rivers in many theoretical studies, exhibit the same connectivity scaling laws as empirical river networks. We found that: i) a subset of connectivity metrics follows clear power-law scaling relationships with network size; ii) different navigation strategies can distort the scaling of some metrics, while others remain unaffected; and iii) scaling relationships in empirical river networks were highly consistent with their OCN-derived counterparts across all metrics, suggesting that OCNs effectively preserve the connectivity properties of real river networks.

\section{Methods}%
\label{sec:methods}
We select a set of 1139 rivers around the world, including the major and most ecologically relevant rivers, with number of nodes ranging from 8 to 1752 (see Fig.~\ref{fig:scheme}a caption for details). This selection correspond to rivers described at resolution level-7 in the HydroSheds database~\cite{hydrosheds}, which guarantees that all connectivity metrics can be computed within feasible limits of time and memory.  At this coarse description, the river basin is partitioned into sub-basins, containing a fragment of the river, following a well established hierarchical method~\cite{hydrosheds}. We, then, construct river networks considering the sub-basins as nodes and edges the connections that enable organisms to move up- and down-stream~(see, for example, Refs.~\cite{muneepeerakul_neutral_2007,muneepeerakul_neutral_2008}).

For each real river network, we then generate a synthetic OCN counterpart with same area $A$ and number of nodes $N$  as follows. We use the recently developed R-package OCNet~\cite{carraro_generation_2020} that first generates synthetic topographies over a $L$-by-$L$ squared grid and then subsequently obtain the river networks via a minimization procedure that assumes a unique outlet~\cite{rinaldo_minimum_1992}. We use grids composed of spatial cells with area, $a$, chosen such that $(aL)^2 = A$. 
Then, for the coarse-graining, we set a minimum sub-basin area, $s_\textrm{min}$, to obtain a network representation of the OCN targeting a specific number of nodes $N$ to also match the real-world network  (see Ref.~\cite{carraro_generation_2020}). To ensure that all sub-basins are represented by many spatial cells of the grid, we set $L=150 \gg \sqrt{A/s_\textrm{min}}$, which more than suffices for all rivers in our dataset (results on larger grids were indistinguishable). With area and number of nodes matched between the datasets, any remaining differences between synthetic and real-world river networks must stem from their connectivity patterns.

With both a real-world river network dataset and a corresponding OCN-based dataset in hand, we then extract size-scaling laws by calculating a set of metrics that capture different facets of connectivity that can arise due the interaction of individual movement behavior and river network structure~\cite{calabrese2004comparison}. In the main text, we focus on the shortest-path harmonic- ($\SH$) and shortest-path betweenness-centrality ($\SB$) metrics, as well as their random-walk variants. Harmonic centrality and betweenness centrality were selected because they provide two complementary connectivity facets, with the former capturing the effective distance between nodes and the latter quantifying the relative importance of nodes in information flows. Briefly, recall that the standard definition of \emph{harmonic centrality}~\cite{MARCHIORI} is given by 
$
    \SH \equiv N^{-1}\sum_{i}^{N} \sum_{j\neq i}^{N} (d_{ij})^{-1}\, ,
$
where $d_{ij}$ is the shortest path (geodesic) between nodes $i$ and $j$. To construct the random-walk harmonic centrality, $\RH$, we replace the shortest distances between a pair of nodes, $d_{ij}$ with the expected random-walk path length that connects them, $\bar d_{ij}$ (see details in Ref.~\cite{noh2004random}). The \emph{betweenness centrality}, $\SB$, of a node is related to the number of shortest paths (between all network node pairs) that cross it~\cite{barrat2004architecture}. 
The random-walk version, known as random-walk betweenness centrality, $\RB$, is obtained by analytically calculating the expected number of simulated random walks that cross a given node in the network~\cite{newman2005measure}, instead of using the shortest paths. A connectivity metric is considered scaling invariant if $\C(\lambda N) \sim \lambda^\beta \C(N)$ holds, i.e if $\C$ is a power-law of the number of nodes, $N$, with a scaling exponent, $\beta$, where the number of nodes serves as an size-related variable in this network perspective.
Finally, our datasets~\cite{RODAREdatasets} and analysis code~\cite{githubColombo} are publicly available (see Sec.~\ref{sec:codedata} for more details).

\section{Results and discussion}

\subsection{Scaling laws for shortest-path based connectivity}
\label{sec:scaling}

\begin{figure}[t!]
    \centering
    \includegraphics[width=0.9\linewidth]{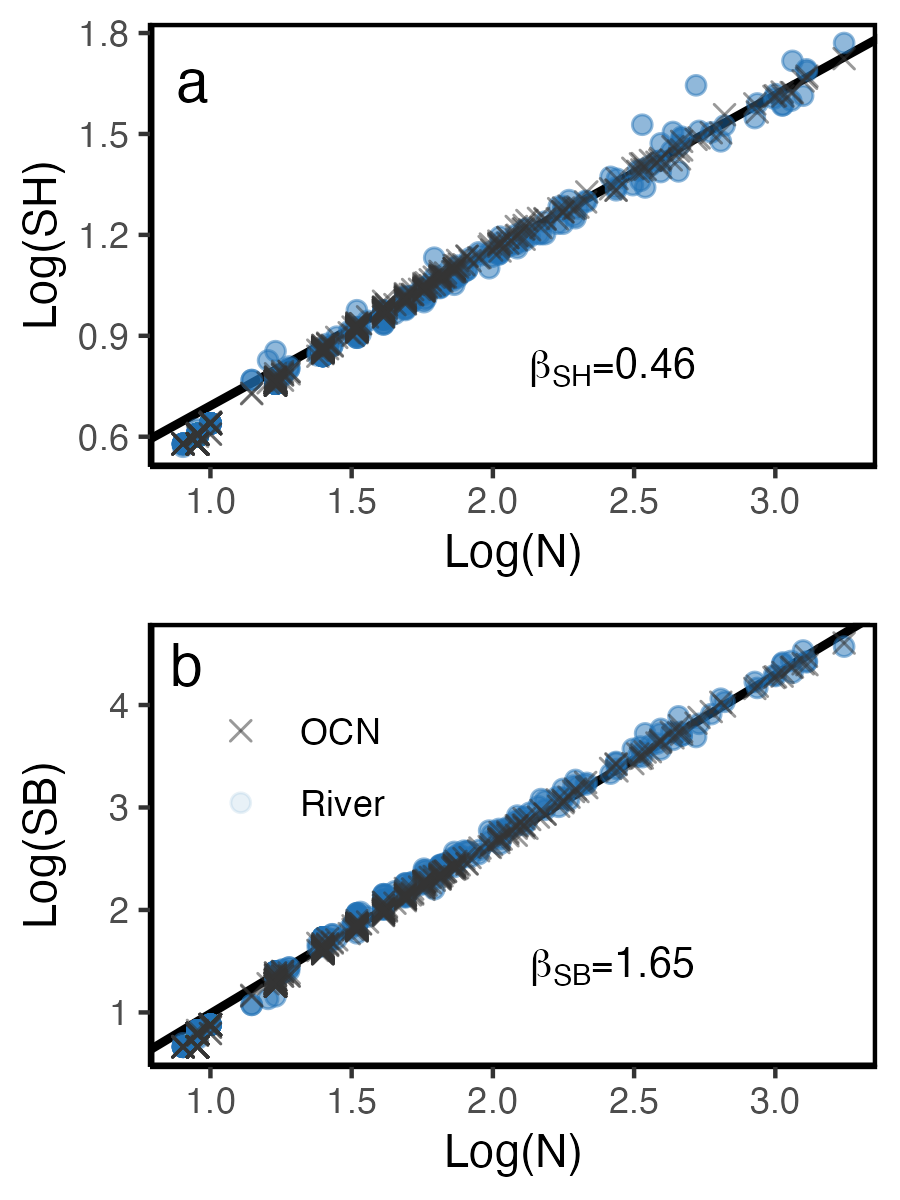}
    \caption{\textbf{Scaling of shortest-path based connectivity metrics.} (a) Harmonic centrality and (b) betweenness centrality as a function of the number of network nodes for real-world rivers (blue circles) and corresponding OCNs (gray crosses). Solid black lines in the background are fits for the OCN networks (with $N>30$) with the slopes, $\beta$, shown in each panel.}
    \label{fig:scaling}
\end{figure}

Both of our focal metrics, $\SH$ and $\SB$, show power-law scaling when plotted against the the number of nodes, $N$ (Fig.~\ref{fig:scaling}). Results for real-world rivers (blue circles) and their synthetic OCN counterparts (black crosses) displayed excellent agreement (Fig.~\ref{fig:scaling}), suggesting that OCNs accurately capture the connectivity patterns of real river networks. In particular, for both datasets, sufficiently large networks (those with at least $\sim$30 nodes) exhibit universal scaling laws, with exponents $0.46$ and $1.64$ (Fig.~\ref{fig:scaling}a,b). 
Examining the broader set of metrics in the \SuppMat~\cite{SM}, we find that closeness centrality, $\SC$, also exhibit clear scale-invariance, which is an expected outcome given that both $\SH$ and $\SC$ are based on the aggregation of the shortest distances between nodes. However, as distances are aggregated differently for $\SH$ and $\SC$, the exponents do not match. In fact, the scaling exponent for $\SC$ is negative, reflecting the nuances of how each metric quantifies centrality (see \SuppMat~for details and Fig.~S1). The remaining metrics, which include degree, robustness, and eigenvector centrality all exhibited either no scaling or weak scaling behavior.

\begin{figure}[t!]
    \centering
    \includegraphics[width=0.9\linewidth]{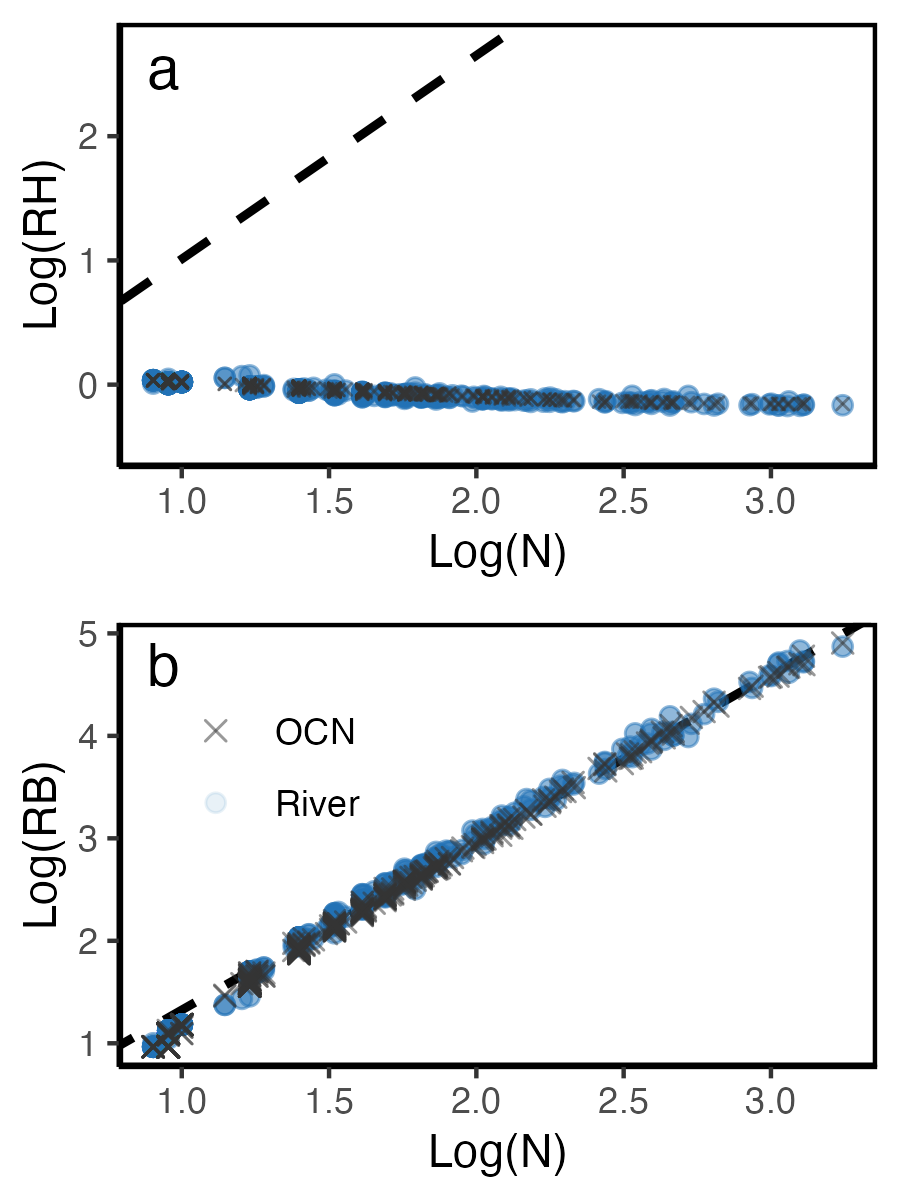}
    \caption{\textbf{Scaling of random-walk based connectivity.} (a) Random-walk harmonic centrality ($\RH$) and (b) Random-walk betweenness centrality ($\RB$) as a function of the number of network nodes for real-world river networks (blue circles) and OCNs (black crosses). Dashed black lines are fits on the log-log scale for the shortest-path based connectivity metrics shown in Fig.~\ref{fig:scaling}a,b.}
    \label{fig:non-scaling}
\end{figure}

\subsection{Scaling laws for random-walk based connectivity}
\label{sec:notscaling}
Focusing first on $\RH$, random navigation does indeed break the scaling law displayed by the standard (shortest-path) $\SH$ metric (cf Fig.~\ref{fig:non-scaling}a to Fig.~\ref{fig:scaling}a). In particular, connectivity goes from a positive to a negative dependence on network size as navigation strategy changes from shortest path to random paths. This drastic change in size dependence arises because random navigation strategies increasingly dilute connectivity as system size increases. Specifically, random navigation explores all possible trajectories between nodes, all but one of which are longer (and some much longer) than the optimal trajectory. The potential for random navigation to explore many wayward paths increases as power-law with network size~\cite{maritan_scaling_1996}, resulting in \emph{lower} connectivity for larger networks. Going from the shortest-path to the random navigation strategies also affected negatively the scaling exponent for the random-walk analogue of the $\SC$ metric, $\RC$. Figure S1 shows that the slope for $\RC$ is significantly steeper than for $\SC$. 

In contrast, for random-walk betweenness centrality (Fig.~\ref{fig:non-scaling}b) the scaling law observed in Fig.~\ref{fig:scaling}b is preserved, suggesting that the $\SB$ connectivity metric is invariant to navigation strategy. This preservation of the scaling law is a product of the unique dendritic structure of rivers, which lack loops, coupled with the definition of betweenness centrality, which counts the net flow through a node~\cite{newman2005measure}. Specifically, under the random navigation strategy, although paths definitely get longer, they do so by entering and returning from dead-end branches. The net flow through these additional visited nodes therefore cancels out, as does their contribution to betweenness~($\RB$). 

\section{Final remarks}

Connectivity, in its many guises~\cite{UrbanKeitt,calabrese2004comparison}, frequently influences the large-scale properties of ecosystems, including biodiversity patterns~\cite{fahrig2003effects} and ecosystem services~\cite{mitchell2015reframing}. The study of connectivity scaling laws in dendritic river networks could, therefore, potentially contribute to our general understanding of the dynamics that generate and maintain riverine biodiversity. Understanding the scaling properties of widely used connectivity metrics on dendritic river networks is therefore an important first step in realizing this promise.

We showed that two widely used network connectivity metrics, $\SH$ and $\SB$, both exhibit power-law scaling with network size when fish are assumed to move optimally via shortest-path navigation. However, the two metrics show dramatically different responses to the opposite assumption of random navigation through the river network. Specifically, the $\RH$ size scaling relationship changes sign relative to $\SH$ while the relationship for $\RB$ is equivalent to that for $\RB$, suggesting invariance to navigation strategy. Other popular network connectivity metrics either showed similar scaling properties to one or both of our focal metrics due to similar mathematical definitions (shortest-path closeness and random-walk closeness), weak scaling with network size (eigenvector-centrality), or no scaling at all (robustness).

The markedly different response of harmonic and betweenness centrality to contrasting navigation assumptions has implications for the use cases of these scaling relationships. Superficially, it would seem that $\SB$ is the metric of choice because it is invariant to movement behavior, which, for many fish species, is difficult to study~\cite{riding2009tracking} and thus poorly known~\cite{radinger2014patterns}. Dendritic systems such as rivers, however, are known to exhibit different biodiversity patterns compared to other geometries such as the 2D lattices that characterize many terrestrial ecosystems. In particular, dendritic systems tend to feature lower local species richness (i.e., lower $\alpha$-diversity) and higher turnover in species composition (i.e., higher $\beta$-diversity) compared to otherwise equivalent 2D lattice systems~\cite{carrara_dendritic_2012}, which may be driven by reduced connectivity. It could therefore be that, for species that lack targeted navigation, $\RH$ captures this potential connectivity limitation better than $\SB$. Further research on the relationships between network structure, functional connectivity, and biodiversity in dendritic systems will be needed to resolve this issue.

Our results also showed that OCNs closely mimic the scaling laws obtained for real-world river networks (Fig.~\ref{fig:scaling},\ref{fig:non-scaling} and Fig.~S1), and that this agreement persists across different navigation strategies. This match between connectivity scaling laws obtained from OCNs and real-world river networks supports earlier claims that OCNs capture the essential statistical properties of river networks ~\cite{maritan_scaling_1996,carraro_optimal_2022}. This result also broadens the use cases for OCNs as mathematically convenient proxies for real river networks to include connectivity and perhaps other large-scale system properties. OCNs are, of course, not exact replicas of empirical river networks, and a more nuanced understanding of how and when OCNs deviate from river networks remains a topic for future study.

\section{Acknowledgments}
This work was partially funded by the Center of Advanced Systems Understanding (CASUS), which is financed by Germany’s Federal Ministry of Education and Research (BMBF) and by the Saxon Ministry for Science, Culture and Tourism (SMWK) with tax funds on the basis of the budget approved by the Saxon State Parliament.

\section{Code and data availability}
\label{sec:codedata}
Code for the generation of the synthetic networks following implementation of Ref.~\cite{carraro_generation_2020} and connectivity metrics calculation can be found at~\cite{githubColombo}. Real-world and OCN river networks and the corresponding metrics are publicly available at \cite{RODAREdatasets}.

\label{sec:data}

\bibliography{scaling_refs}

\end{document}